\documentclass[prl,twocolumn,showpacs,superscriptaddress,nofootinbib]{revtex4-1}
\usepackage{amsmath}
\usepackage{dcolumn}
\usepackage{bm}
\usepackage{graphicx}
\usepackage{amsfonts}
\usepackage{amssymb}
\usepackage{amstext}
\usepackage{bbm}
\usepackage{dsfont}
\usepackage{float}
\usepackage[usenames,dvipsnames]{xcolor}
\usepackage[colorlinks=true,citecolor=Cerulean,linkcolor=RubineRed,urlcolor=Cerulean]{hyperref}

\newcommand{\beq}{\begin{equation}}
\newcommand{\eeq}{\end{equation}}
\newcommand{\beqa}{\begin{eqnarray}}
\newcommand{\eeqa}{\end{eqnarray}}

\begin{document}

\title{Extreme Decoherence and Quantum Chaos}
\author{Zhenyu Xu}
\affiliation{School of Physical Science and Technology, Soochow University, Suzhou 215006, China}
\affiliation{Department of Physics, University of Massachusetts, Boston, MA 02125, USA}

\author{Luis Pedro Garc\'ia-Pintos}
\affiliation{Department of Physics, University of Massachusetts, Boston, MA 02125, USA}

\author{Aur\'elia Chenu}
\affiliation{Donostia International Physics Center,  E-20018 San Sebasti\'an, Spain}
\affiliation{IKERBASQUE, Basque Foundation for Science, E-48013 Bilbao, Spain}
\affiliation{Theoretical Division, Los Alamos National Laboratory, MS-B213, Los Alamos, NM 87545, USA}

\author{Adolfo del Campo}
\affiliation{Donostia International Physics Center,  E-20018 San Sebasti\'an, Spain}
\affiliation{IKERBASQUE, Basque Foundation for Science, E-48013 Bilbao, Spain}
\affiliation{Department of Physics, University of Massachusetts, Boston, MA 02125, USA}
\affiliation{Theoretical Division, Los Alamos National Laboratory, MS-B213, Los Alamos, NM 87545, USA}

\begin{abstract}
We study the ultimate limits to the decoherence rate associated with
dephasing processes. Fluctuating chaotic quantum systems are shown to
exhibit extreme decoherence, with a rate that scales exponentially with the
particle number, thus exceeding the polynomial dependence of systems with
fluctuating $k$-body interactions. Our findings suggest the use of quantum chaotic systems as a natural test-bed for spontaneous wave function collapse models.
We further discuss the implications on the decoherence of AdS/CFT black holes resulting from the  unitarity loss  associated with energy dephasing.
\end{abstract}

\maketitle

Decoherence is a ubiquitous phenomenon in nature, that is responsible for
the emergence of classical behavior from the quantum substrate \cite%
{Rev03,Rev05,Book-Decoherence}. Different sources of decoherence can be
identified. Decoherence is most commonly attributed to the interaction
between the system and its surrounding environment. However, it can also
arise from the presence of random fluctuations in the system evolution.
These can have an intrinsic quantum origin, as in the case of continuously
monitored systems, or be associated with classical sources of noise, as
those described by fluctuating Hamiltonians. In each case the dynamics
becomes stochastic, and upon averaging over realizations of the noise
processes, decoherence manifests itself in an ensemble perspective. This
scenario has important applications in quantum optics \cite{Milburn98},
quantum simulations \cite{Rev12,Rev14,Dutta2016,Chenu2017}, quantum sensing
\cite{sensing}, and collapse models \cite{Bassi13}. In addition, the
decoherence effect can also be independent of any direct interaction with an
environment or noise fluctuations. In particular, a loss of unitarity can
arise when quantum dynamics exhibits random phase changes on a short time
scale~\cite{Milburn91}, in the description of quantum evolution with
realistic clocks of finite precision~\cite{clock,Diosi,gravity}, or as a
result of gravitational effects \cite{Blencowe}.

In general, decoherence increases with the system size \cite{Rev05}, making
challenging quantum information and simulation tasks with complex quantum
systems involving a large number of particles and degrees of freedom \cite%
{Rev08,Rev17}. For a complex quantum system that exhibits chaos, decoherence
can be expected to be singular due to the enhanced sensitivity to initial
conditions. Chaotic quantum systems can be described using random matrix
Hamiltonians with appropriate symmetries \cite{Book-RMT,Book-RMT1}.
Originally, random matrix theory was introduced by Wigner to deal with the
statistics of the spectra of heavy atomic nuclei \cite%
{Book-RMT,Book-RMT1,Book-RMT2,Rev-nuclear,Rev-nuclear2}.\ Recent progress includes
applications to complex open quantum systems \cite{Mata11,Breuer11,Breuer13}%
, Majorana fermions \cite{Rev-F}, many-body quantum chaos \cite{Kos18,Chan},
work statistics in chaotic systems \cite{Chenu17,Chenu18,work}, and
information scrambling in black holes \cite%
{Preskill,scramble,Barbon,Maldacena2016,Dyer17,Cotler1,delCampo17prd,Cotler2,Vegh}%
.

Understanding the interplay between quantum chaos and decoherence is a longstanding problem. Earlier studies of this subject are mainly focused on the effect of dissipation and decoherence on level statistics, and how to incorporate such effects into the study of quantum systems that are chaotic in the classical limit \cite{Book-RMT1,Book-Chaos2}. Here, we pose the question as to what is the ultimate limit to the rate of
decoherence of complex quantum systems. This issue is not only of relevance
to fundamental and applied aspects of quantum science and technology, but
has implications that extend to other fields, including black-hole physics.
In this Letter, we introduce a decoherence rate that applies to arbitrary
Markovian processes. Using it, we show that the dynamics of fluctuating
chaotic quantum systems is extreme in that its rate scales exponentially
with the number of particles $n$. Such scaling has no match in physical
systems with $k$-body interactions, where the decoherence rate scales
polynomially with $n$.  In turn, this allows us to identify chaotic quantum systems described
by random matrix theory as a natural test-bed for spontaneous wavefunction collapse models.

\textit{Decoherence rates.---} A Markovian open quantum system is generally
described by a master equation of the Lindblad form \cite{Lindblad,Book-Open}
\begin{equation}
\dot{\rho}_{t}=-\frac{i}{\hbar }\left[ H,\rho _{t}\right] +\sum_{\mu }\gamma
_{\mu }\left( V_{\mu }\rho _{t}V_{\mu }^{\dag }-\frac{1}{2}\{V_{\mu }^{\dag
}V_{\mu },\rho _{t}\}\right) ,  \label{Lindblad}
\end{equation}%
where the Hamiltonian $H$ determines the unitary evolution of the system,
the coupling constants $\gamma _{\mu }\geq 0$ are non-negative, and $V_{\mu
} $ are the corresponding Lindblad operators.

In order to characterize how fast the system decoheres, we consider the
purity $P_{t}:=\mathrm{{tr}\left( \rho _{t}^{2}\right) }$ of the state,
which quantifies its degree of mixedness. An expression for the decoherence
rate under Markovian evolution can be obtained from the short-time
asymptotic behavior
$P_{t}=$tr$\left( \rho _{0}^{2}\right) +2$tr$\left( \rho _{0}\dot{\rho}%
_{0}\right) t+\mathcal{O}(t^{2})\simeq P_{0}(1-Dt)$, from which we define
the decoherence rate \cite{note}
\begin{equation}
D:=-\frac{2\text{tr}\left( \rho _{0}\dot{\rho}_{0}\right) }{\text{tr}\left(
\rho _{0}^{2}\right) }=\frac{2\sum_{\mu }\gamma _{\mu }\widetilde{\mathrm{cov%
}}_{\rho _{0}}(V_{\mu }^{\dag },V_{\mu })}{\text{tr}\left( \rho
_{0}^{2}\right) },  \label{DR}
\end{equation}%
where $\widetilde{\mathrm{cov}}_{\rho _{0}}(X,Y):=\left\langle \rho
_{0}XY\right\rangle _{\rho _{0}}-\left\langle X\rho _{0}Y\right\rangle
_{\rho _{0}}$ is the modified covariance, with $\left\langle X\right\rangle
_{\rho _{0}}:=$tr$\left( \rho _{0}X\right) $. Equation (\ref{DR}) depends
only on the initial state and Lindblad operators, facilitating the analysis
of decoherence in complex quantum systems without the need to solve the
dynamical equations of motion. Equation (\ref{DR}) further extends Zurek's
seminal estimate of the decoherence time \cite{Rev03}, derived in the
context of quantum Brownian motion, to arbitrary Markovian dynamics.

In what follows we shall be interested in Hermitian Lindblad operators $%
V_{\mu }$, when Eq.~\eqref{Lindblad} can also be rewritten in a
double-commutator form. The purity is then guaranteed to decrease
monotonically with time, $\dot{P}_{t}<0$ for all $t\geq0$ \cite{Lidar06}. In
particular, we shall focus on $V_{\mu }$ described by random matrices as
well as $k$-body interaction operators. Such instances of $V_{\mu }$ are natural in a
wide variety of scenarios including the description of fluctuating
Hamiltonian systems and collapse models, as described below.

\textit{Extreme decoherence rates with random-matrix Lindblad operators.---}
Statistical spectral properties of quantum chaotic systems can be
conveniently described using ensembles of random matrix operators $\{X\}$
\cite{Book-RMT,Book-RMT1}. Gaussian ensembles are associated with operators
in which matrix elements are i.i.d. complex Gaussian variables. Gaussian
ensembles can be classified in terms of the invariance of the joint
eigenvalue density $\varrho (X)$ under similarity transforms, $\varrho
(X)=\varrho (UXU^{-1})$. Prominent examples include orthogonal, unitary or
symplectic matrices $U$ \cite{Book-RMT2}. When the dimension of the Hilbert
space is large, properties such as the eigenvalue density become universal
and are shared by the different ensembles.

We shall consider the decoherence under chaotic Lindblad operators $V_{\mu }$
sampled from the Gaussian Unitary Ensemble (\textrm{GUE}) with dimension $d$%
, i.e., $V_{\mu }\in $ \textrm{GUE}. To this end, we introduce a simplified
\textrm{GUE} average of a function $f(X)$ ($X\in $ \textrm{GUE}) with Haar
measure \cite{SM}
\begin{equation}
\left\langle f(X)\right\rangle _{\mathrm{GUE}}:=\int \prod_{k=1}^{d}\mathrm{d%
}x_{k}\varrho _{\mathrm{GUE}}(x_{1},\dots ,x_{d})\left\langle
f(X)\right\rangle _{\text{\textrm{Haar}}},  \label{ave}
\end{equation}%
where $\varrho _{\mathrm{GUE}}(x_{1},\dots ,x_{d})$ is the $d$-point
correlation function for the eigenvalues \{$x_{k}$\}$_{k=1,\cdots ,d}$ of $X$%
, and $\left\langle f(X)\right\rangle _{\text{\textrm{Haar}}}:=\int_{%
\mathcal{U}(d)}f(UHU^{-1})\mathbf{d}\mu (U)$ denotes the Haar average over
the unitary group $\mathcal{U}(d)$ with Haar measure $\mathbf{d}\mu (U)$
\cite{Book-Haar,Haar,Book-RMT3,Book-Tao}. Without loss of generality, the
initial state is assumed to be pure with $\rho _{0}=\left\vert \Psi
_{0}\right\rangle \left\langle \Psi _{0}\right\vert $ (see \cite{SM} for the
mixed state case). As shown in \cite{SM}, when the initial pure state $\rho
_{0}$ is fixed and chosen independently of $V_{\mu }$, the decoherence rate
with averaged over GUE reads
\begin{equation}
D_{\mathrm{GUE}}=\frac{2d}{d+1}\sum_{\mu }\gamma _{\mu }\langle \mathrm{var}%
_{\rho _{\beta =0}}(V_{\mu })\rangle _{\mathrm{GUE}}=\frac{\Gamma d^{2}}{d+1}%
,  \label{DR Haar}
\end{equation}%
where $\mathrm{var}_{_{\rho }}(X):=\left\langle X^{2}\right\rangle _{\rho
}-\left\langle X\right\rangle _{\rho }^{2}$ denotes the variance of $X$ in
the state $\rho $, $\rho _{\beta =0}=\mathds{1}_{d}/d$ is the thermal state
at infinite temperature, and $\Gamma :=\sum_{\mu }\gamma _{\mu }$. Note that
information regarding the initial state is lost with the Haar average in the
first equality, as the variance is expressed only in terms of the thermal
state at infinite temperature. In Eq. (\ref{DR Haar}), the second equality
follows from using the corresponding correlation function $\varrho _{\mathrm{%
GUE}}$ of GUE. We note that the scaling of the decoherence rate in Eq. (\ref%
{DR Haar}) stems from the dependence of the density of states on the Hilbert
space dimension in systems described by random matrices. In particular, it
is independent of other spectral signatures of chaos, such as the level
spacing distribution \cite{Book-RMT,Book-RMT1}.

\begin{figure}[tbp]
\centering{}\includegraphics[width=3.25in]{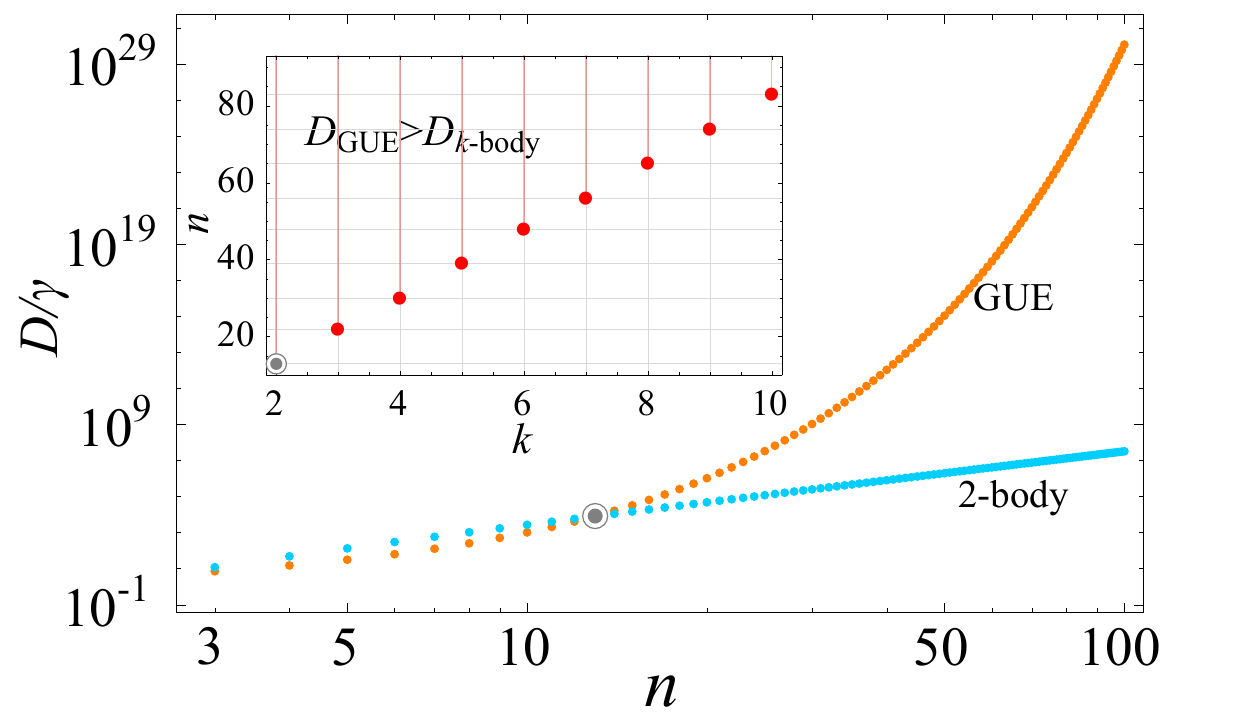}
\caption{\textbf{Extreme decoherence.} Comparisson of the decoherence rates
associated with fluctuating chaotic systems $D_{\mathrm{GUE}}$ and $k$-body
systems $D_{k-\mathrm{body}}$, with $k=2$ as an example. Inset: The red dots
represent the minimum particle number $n$ for which the decoherence rate in
a chaotic system surpasses that of $k$-body system. The condition when $D_{%
\mathrm{GUE}}>$ $D_{k-\mathrm{body}}$ is $k\lesssim [n/10+1]$.}
\label{Fig_extreme}
\end{figure}

For simplicity, we consider a single chaotic operator $V$ with rate $\gamma $%
. Assuming for the sake of illustration that the system is composed of $n$
qubits with a Hilbert space dimension $d=2^{n}$, the corresponding
decoherence rate for chaotic operators becomes extremely fast, with $D_{%
\mathrm{GUE}}=\gamma 2^{2n}/(2^{n}+1)\simeq \gamma 2^{n}$. Decoherence is
then exponentially faster than in the case of $k$-body Lindblad operators,
except for extremely non-local interactions, with $k\gtrsim \mathcal{O}(n)$.
To show this, consider the general case of a $k$-body Lindblad operator of
the form
\begin{equation}
V=\epsilon \sum_{l_{1}<\cdots <l_{k}}^{n}\Lambda _{l_{1}<\cdots <l_{k}},
\label{k-body}
\end{equation}%
where $\epsilon $ is a dimensionless positive constant to be determined by
comparison with GUE, as discussed below. The variance is bounded by $\mathrm{%
var}_{_{\rho _{0}}}(V)\leq \left\Vert V\right\Vert ^{2}$, where $\left\Vert
X\right\Vert :=x_{M}$ is the spectral norm and $x_{M}$ is the maximum
eigenvalue of $\sqrt{X^{\dag }X}$ \cite{SM}. The spectral norm of $V$ in Eq.
(\ref{k-body}) is given by
\begin{equation}
\left\Vert V\right\Vert \leq \epsilon \sum_{l_{1}<\cdots
<l_{k}}^{n}\left\Vert \Lambda _{l_{1}<\cdots <l_{k}}\right\Vert =\epsilon
\left\Vert \Lambda _{l_{1}<\cdots <l_{k}}\right\Vert \binom{n}{k},
\end{equation}%
where $\binom{n}{k}:=\frac{n!}{k!(n-k)!}$ is the binomial coefficient.
Therefore, the decoherence rate for the $k$-body case satisfies%
\begin{equation}
D_{k-\mathrm{body}}\lesssim \frac{2\gamma \epsilon^{2} \left\Vert \Lambda
_{l_{1}<\cdots <l_{k}}\right\Vert ^{2}}{(k!)^{2}}n^{2k},  \label{DR k-body}
\end{equation}%
where we have assumed $k\ll n$. Said differently, the decoherence rate of
the $k$-body system grows at most polynomially in $n$. For the sake of
illustration, we consider an example in which the decoherence operator given
by a $k$-body all-to-all long-range term $\Lambda _{l_{1}<\cdots
<l_{k}}=\sigma _{l_{1}}^{z}\otimes \cdots \otimes \sigma _{l_{k}}^{z}\otimes %
\mathds{1}_{\delta \neq l_{1},\cdots ,l_{k}}$, where $\sigma ^{z}$ is the
usual Pauli operator. As $\left\Vert \Lambda _{l_{1}<\cdots
<l_{k}}\right\Vert \equiv 1$, we have $D_{k-\mathrm{body}}\lesssim 2\gamma
\epsilon^{2} n^{2k}/(k!)^{2}$. In order to perform a direct comparison
between GUE and $k$-body systems, we set $D_{\text{GUE}}(n_{0})=D_{k\text{%
-body}}(n_{0})$, where $n_{0}$ is an arbitrary starting reference point for
the particle number, with which the parameter $\epsilon $ can be determined.
Fig.~\ref{Fig_extreme} presents numerical calculations for $n_{0}=1 $ and $%
\epsilon^2 =2/3$ as an example, showing that the decoherence rate for random
operators is larger than for $k$ -body ones as long as $k\lesssim [n/10+1]$ in
this case. This implies that chaotic dephasing described by random matrix
theory not only leads to decoherence in an extreme way, but also faster than
the physical $k$-body quantum systems in high dimensional situations, which
we illustrate in Fig.~\ref{Fig_extreme}.
We expand this discussion with a spin model of two-body random ensembles \cite{TBRE} in the Supplemental Material \cite{SM}.

\textit{Decoherence rates of entangled states.---} In what follows we
illustrate extreme decoherence in an entangled state. For simplicity, we consider a thermal
state of the form $\rho_{0}=e^{-\beta H}/Z(\beta )$, where the
normalization constant is given by the partition function $Z(\beta ):=%
\mathrm{tr}(e^{-\beta H})$, with inverse temperature $\beta =(k_{B}T)^{-1}$.
This is a mixed state that can be purified by doubling the Hilbert space
dimension and considering two identical copies of the system. The resulting
entangled state is known as the thermofield double (TFD) state \cite%
{Book-TFD}%
\begin{equation}
\left\vert \Phi _{0}\right\rangle :=\frac{1}{\sqrt{Z(\beta )}}\sum_{k}e^{-%
\frac{\beta E_{k}}{2}}\left\vert k\right\rangle \left\vert k\right\rangle .
\label{TDS}
\end{equation}%
where $E_{k}$ ($\left\vert k\right\rangle $) are the corresponding
eigenvalues (eigenvectors) of $H$. Tracing over any of the subsystems
recovers the thermal state. TFD states are commonly used in
finite-temperature field theory and have been widely studied in the context
of holography, e.g., in connection to the entanglement between black holes~\cite{Maldacena2013}, the butterfly effect~ \cite{butterfly}, and quantum
source-channel codes~\cite{code}.

Here, we focus on sources of decoherence that act on the energy basis, as
those arising in certain spontaneous wave function collapse models,
that constitute stochastic modifications of quantum mechanics leading to localization in energy  \cite{Gisin84,Percival94,SSE-PRD,Bassi13,BG03}.
An equivalent source of decoherence arises in the presence of
fluctuating fields or coupling constants in the Hamiltonian \cite{Chenu2017} and random measurement Hamiltonians \cite{Korbicz17}.
As already noted, decoherence in the energy eigenbasis arises as well if
quantum dynamics at short time scales includes intrinsic uncertainties \cite%
{Milburn91} or when the time-evolution is described according to a realistic
clock of finite precision; see \cite{clock,gravity} and \cite{SM}.

Decoherence in the energy eigenbasis can generally be described in terms of
fluctuating Hamiltonians. Assume now that each subsystem is perturbed by a
single Gaussian real white noise $\xi _{t}$ \cite{Book-Stochastic}, i.e., $%
H\rightarrow (1+\hbar \sqrt{\gamma _{L(R)}}\xi _{t}^{L(R)})H$ when $t>0$.
Then, the total Hamiltonian is given by $\tilde{H}_{t}=H\otimes \mathds{1}+%
\mathds{1}\otimes H+\hbar \sqrt{\gamma }(\xi _{t}^{L}H\otimes \mathds{1}+%
\mathds{1}\otimes \xi _{t}^{R}H)$, where we assume independent noises $\xi
_{t}^{L}\neq \xi _{t}^{R}$, with identical amplitudes $\gamma _{L}=\gamma
_{R}=\gamma $ (here the tilde \textquotedblleft $%
\sim $\textquotedblright\ is used to represent the global Hilbert space of
the two copies). The dynamics of the system is governed by the Schr\"{o}dinger
equation $i\hbar \partial _{t}|\Psi _{t}^{\xi }\rangle =\tilde{H}_{t}|\Psi
_{t}^{\xi }\rangle $, denoting $\xi :=\{\xi _{t}^{L},\xi _{t}^{R}\} $ for
simplicity. Using Novikov's theorem~\cite{Novikov} or It\^{o} calculus~\cite%
{SSE-PRD}, one can show that the dynamics of the noise-averaged density
matrix $\rho _{t}=\langle |\Psi _{t}^{\xi }\rangle \langle \Psi _{t}^{\xi
}|\rangle _{\xi }$ is governed by the master equation~\eqref{Lindblad}, with
Lindblad operators $\tilde{V}_{1}=H\otimes \mathds{1}$ and $\tilde{V}_{2}=%
\mathds{1}\otimes H$; see details in \cite{SM}. The exact evolution of the
density matrix reads \cite{SM},
\begin{equation}
\rho _{t}=\frac{1}{Z(\beta )}\sum_{k,\ell }e^{-\frac{\beta }{2}E_{k\ell
}^{+}-i\frac{2t}{\hbar }E_{k\ell }^{-}-\gamma t(E_{k\ell
}^{-})^{2}}|k\rangle |k\rangle \langle \ell |\langle \ell |,
\end{equation}%
where $E_{k\ell }^{\pm }:=E_{k}\pm E_{\ell }$. The decay of the purity is
thus given by
\begin{equation}
P_{t}=\sqrt{\frac{1}{8\pi \gamma t}}\int_{-\infty }^{\infty }e^{-\frac{y^{2}%
}{8\gamma t}}\left\vert \frac{Z(\beta -iy)}{Z(\beta )}\right\vert ^{2}%
\mathrm{d}y,
\end{equation}%
in terms of the analytic continuation of the partition function. At long
times, the purity approaches $P_{\infty }=Z(2\beta )/Z(\beta )^{2}$, which
is the purity of a canonical thermal state at temperature $\beta $ and
reduces to $1/d$ at infinite temperature ($\beta =0$). The corresponding
decoherence rate can be immediately obtained from Eq.~\eqref{DR}
\begin{equation}
\tilde{D}=4\gamma \mathrm{var}_{\rho _{\beta }}(H)=4\gamma \frac{\mathrm{d}%
^{2}}{\mathrm{d}\beta ^{2}}\ln \left[ Z(\beta )\right] ,  \label{DR TDS}
\end{equation}%
where $Z(\beta )=\int \mathrm{d}E\varrho (E)e^{-\beta E}$, and $\varrho (E)$
is the spectral density of $H$.

\begin{figure}[tbp]
\centering{}\includegraphics[width=3.25in]{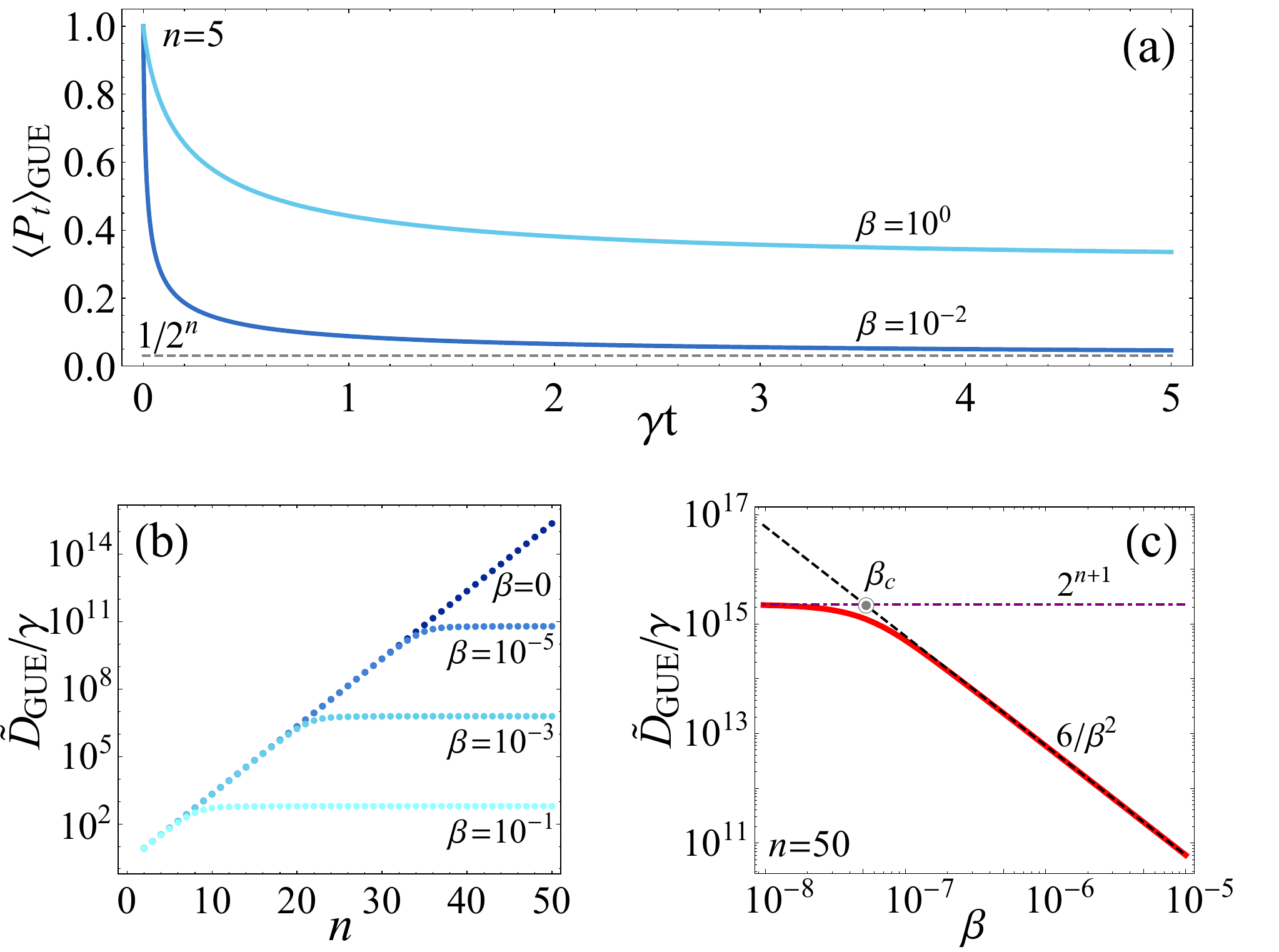}
\caption{\textbf{Decoherence of a thermofield double state.} (a) Purity $%
\left\langle P_{t}\right\rangle _{\text{\textrm{GUE}}}$ as a function of $%
\protect\gamma t$ with different temperatures $\protect\beta$. The numerical
average is performed over 1000 realizations of GUE with $n= 5$. (b) The
decoherence rates for GUE $\tilde{D}_{\mathrm{GUE}}/\protect\gamma $ versus
the particle (treated as qubits) number $n$ for different temperatures $%
\protect\beta$. With the increase in the Hilbert space dimension, $\tilde{D}%
_{\mathrm{GUE}}/\protect\gamma $ saturates at $6/\protect\beta^{2}$. (c) The
decoherence rates (red solid curve) for GUE $\tilde{D}_{\mathrm{GUE}}/%
\protect\gamma $ as a function of $\protect\beta$ with $n=50$. The
asymptotic expressions for $\tilde{D}_{\mathrm{GUE}}/\protect\gamma =2^{n+1}$
under $\protect\beta\ll \protect\beta _{c}:=\protect\sqrt{3/2^{n}}$ (purple
dotted-dashed line) and $\tilde{D}_{\mathrm{GUE}}/\protect\gamma =6/\protect%
\beta^{2}$ under $\protect\beta\gg \protect\beta _{c}$ (black dashed line)
are also shown.}
\label{Fig_TDS}
\end{figure}

Let us first consider $H$ modeled by a random matrix sampled from GUE \cite%
{Maldacena2016,Cotler1}. Given that the initial TFD state is defined in
terms of the Hamiltonian $H$, some care is needed when performing the
average in this case. In the following, we calculate the purity $%
\left\langle P_{t}\right\rangle _{\text{\textrm{GUE}}}$ shown in Fig. \ref%
{Fig_TDS}(a), and characterize the decoherence rate $\tilde{D}_{\text{%
\textrm{GUE}}}$. From Eq.~\eqref{DR
TDS}, the latter is given by
\begin{equation}
\tilde{D}_{\text{\textrm{GUE}}}\doteq 4\gamma \frac{\mathrm{d}^{2}}{\mathrm{d%
}\beta ^{2}}\ln [\left\langle Z(\beta )\right\rangle _{\text{\textrm{GUE}}}],
\label{DR TDS GUE}
\end{equation}%
where \textquotedblleft $\doteq $\textquotedblright\ indicates the use of
the annealing approximation, that we show to be highly accurate in \cite{SM}%
. For the average $\left\langle Z(\beta )\right\rangle _{\text{\textrm{GUE}}%
}=\int \mathrm{d}E\varrho _{\text{\textrm{GUE}}}(E)e^{-\beta E}$, using
Wigner's semicircle law in the limit of $d\gg 1$ one finds $\left\langle
Z(\beta )\right\rangle _{\text{\textrm{GUE}}}=\sqrt{2d}\,\mathrm{I}_{1}(%
\sqrt{2d}\beta )/\beta $, where $\mathrm{I}_{n}(\cdot )$ is the modified
Bessel function of first kind and order $n$ \cite{Book-Int}. From Eq. (\ref%
{DR TDS GUE}), it then follows that \cite{SM}
\begin{equation}
\tilde{D}_{\mathrm{GUE}}=8\gamma d\left[ 1-\frac{3}{\sqrt{2d}\beta }g(\sqrt{%
2d}\beta )-g(\sqrt{2d}\beta )^{2}\right] ,  \label{F3}
\end{equation}%
with $g(x):=\mathrm{I}_{2}(x)/\mathrm{I}_{1}(x)$. The decoherence rate is
depicted as a function of particle number under different $\beta $ in Fig. %
\ref{Fig_TDS}(b). Specifically, as a function of the temperature, the
decoherence rate in Eq. (\ref{F3}) simplifies to
\begin{equation}
{\tilde{D}}_{\text{\textrm{GUE}}}\simeq \left\{
\begin{array}{ll}
2\gamma d, & (\beta \ll \beta _{c}), \\
\frac{6\gamma }{\beta ^{2}}, & (\beta \gg \beta _{c}),%
\end{array}%
\right.   \label{limit}
\end{equation}%
where $\beta _{c}:=\sqrt{3/d}$. In the high temperature limit, when $\beta
\ll $ $\beta _{c}$ the decoherence time reduces to ${\tilde{D}}_{\text{%
\textrm{GUE}}}\simeq 2\gamma d$ in agreement with Eq. (\ref{DR Haar}). When $%
\beta \gg $ $\beta _{c}$, ${\tilde{D}}_{\text{\textrm{GUE}}}$ asymptotically
approaches to $6\gamma /\beta ^{2},$ proportional to the
temperature square, i.e., at low temperature decoherence is highly
suppressed, as shown in Fig. \ref{Fig_TDS}(c).

Next we illustrate the extent to which the decoherence of fluctuating
quantum chaotic Hamiltonians is extreme and faster than physical systems
with 2-body interactions. To this end, we compare the decoherence rate of a
high-temperature TFD state of a chaotic quantum system with that of a spin
Hamiltonian with all-to-all long-range $2$-body interactions. In particular,
we consider Lipkin-Meshkov-Glick model with zero external magnetic field,
with a Hamiltonian $H_{0}=\epsilon \sum_{l<m}^{n}\sigma _{l}^{z}\otimes
\sigma _{m}^{z}$ ($n\geq 2$) that is amenable to quantum simulation \cite%
{XPeng,Rey}. For the latter, in the high temperature limit $\tilde{D}_{\text{%
\textrm{2-body}}}|_{\beta\ll \beta _{c}}\simeq 2\gamma n(n-1)$, which has a
polynomial dependence on $n$. In spite of the all-to-all pairwise
interactions, the rate is slower than that in the GUE case, that is
characterized by extreme decoherence ($\tilde{D}_{\text{\textrm{GUE}}%
}|_{\beta\ll \beta _{c}}\simeq 2\gamma 2^{n}$ for qubits).

\textit{Decohrence of AdS/CFT black holes.---}
The decoherence of black holes induced by Hawking radiation has been widely
studied~\cite{Rev-BH,Demers,Arrasmith}.
The preceding analysis can  be  applied to the decoherence of eternal AdS-Schwarzschild
black holes. According to the AdS/CFT correspondence, quantum gravity in an
asymptotically anti-de Sitter (AdS) space-time is dual to a
non-gravitational conformal field theory (CFT) on a lower dimensional
space-time \cite{AdS,Rev-BH,Mark}.
Here, we consider
a thermofield double (TFD) state of two non-interacting copies of CFT \cite%
{TFD,Maldacena2003,rev}. This can be interpreted as two entangled black
holes in disconnected space, with common time. EPR correlations in the TFD
state make the geometry of the two black holes connected by an Einstein-Rosen
(ER) bridge.
Said differently, one possible interpretation of the TFD
state is that it is dual to two eternal AdS-Schwarzschild black holes in
disconnected spaces with a common time \cite{Maldacena2013}. In this case,
the total Hamiltonian is simply the sum of the two CFT Hamiltonians $\tilde{H%
}=H\otimes \mathds{1}+\mathds{1}\otimes H$ in Hilbert spaces $\mathcal{H}%
\otimes \mathcal{H}$ \cite{Maldacena2013}. According to the EPR=ER conjecture \cite{Maldacena2013,EREPR1,EREPR2}, the
presence of entanglement and EPR correlations is associated with a geometry
of an Einstein-Rosen bridge describing the entangled black holes.
Unitarity loss associated with energy dephasing leads to the decay of of quantum correlations.
The resulting decoherence of the TFD state results in the  closing of the
Einstein-Rosen bridge. While certain aspects of the late-time behavior of AdS/CFT black holes are captured by random matrix theory \cite{Cotler1}, decoherence is however not extreme in this context.
Indeed, the decoherence rate can be written in terms of the heat capacity $C$ of the CFT as $\tilde{D}=4\gamma C/(k_B\beta^2)$.
The latter is proportional to the the entropy of the black hole $S$, which scales  with the number of degrees of freedom $n$ \cite{PR15}.

\textit{Discussion and Conclusions.---} We have introduced a decoherence
rate for arbitrary Markovian processes and used it to demonstrate that
fluctuating chaotic systems described by random matrix theory exhibit
extreme decoherence. The latter is characterized by a rate that grows
exponentially with the particle number, thus surpassing the dynamics of
non-chaotic $k$-body Hamiltonians. This conclusion holds generally for any
source of decoherence acting on the energy eigenbasis.
Our findings suggest
that chaotic quantum systems provide an ideal test-bed to explore deviations
from quantum mechanics, such as those predicted by spontaneous wave function collapse models.
This identification motivates the extension of current experimental efforts \cite{Bassi16,Bassi17} to probe decoherence  in the energy basis.
We have also applied  our analysis to the fate of black holes under unitarity loss in the context of AdS/CFT and shown  that the decoherence is not extreme in this context, in spite of the known random-matrix behavior at long times.

A surge of activity has recently been devoted to probing aspects of many-body quantum chaos and related models in a variety of platforms including trapped ions \cite{Rey,ions2}, nuclear
magnetic resonance systems \cite{Du,Luo17}, ultracold atoms \cite{ultracold atom1,ultracold atom2,ultracold atom3,tocape}, and superconducting qubits \cite%
{Laura17}. In particular, the generation of Haar-uniform random operations has been proposed in many-body systems driven by stochastic external pulses \cite{Banchi17}.
We hope that the present work stimulates both theoretical and
experimental research on the extreme decoherence rates in chaotic complex
quantum systems, that is at reach with current technology.

\textit{Acknowledgements}.---The authors are indebted  to B. Swingle for clarifications on black hole decoherence. It is also a pleasure to thank F. J. G\'omez-Ruiz,
J. Molina-Vilaplana, J. Sonner and J. Maldacena for  feedback on the manuscript. We acknowledge
funding support by UMass Boston (project P20150000029279), the John
Templeton Foundation, and the National Natural Science Foundation of China
(Grant No. 11674238).

\pagebreak
\widetext
\begin{center}
\vspace{16cm}
\textbf{{\large Supplemental Material--Extreme Decoherence: from Quantum Chaos to Black Holes}}
\end{center}

\setcounter{equation}{0} \setcounter{figure}{0} \setcounter{table}{0}
\makeatletter
\renewcommand{\theequation}{S\arabic{equation}} \renewcommand{\thefigure}{SM%
\arabic{figure}} \renewcommand{\bibnumfmt}[1]{[#1]} \renewcommand{%
\citenumfont}[1]{#1}

\tableofcontents

\subsection{A. Derivation of decoherence rate in GUE}

\subsubsection{\textit{1. GUE average and Haar measure}}

The \textrm{GUE }average of function $f(X)$ ($X\in $\textrm{GUE }with
dimension $d$), denoted as
\begin{equation}
\langle f(X)\rangle _{\mathrm{GUE}}:=\int f(X)\mathbf{d}\mu (X),
\label{GUE1}
\end{equation}%
is obtained for the ensemble probability measure
\begin{equation}
\mathbf{d}\mu (X):=Ce^{-\mathrm{{tr}(}X^{2}\mathrm{)}}\mathbf{d}X,\text{
with }\mathbf{d}X:=\prod_{j}\mathrm{d}X_{jj}\prod_{k<l}\mathrm{d}(\mathrm{Re}%
X_{kl})\,\mathrm{d}(\mathrm{Im}X_{kl}),  \label{GUE2}
\end{equation}%
where $C$ is a normalization constant, given in e.g. Ref. \cite{sBook-RMT},
and $\mathbf{d}X$ is the flat Lebesgue measure on $d\times d$ Hermitian
matrices $X$. Every $X\in $\textrm{GUE} can be diagonalized with a unitary
operator $U$ [$U\in \mathcal{U}(d)$], i.e., $X=U\tilde{X}U^{-1}$, with $%
\tilde{X}_{jk}=x_{j}\delta _{jk}$. Thus,
\begin{equation}
\mathrm{d}X:=\left( \mathrm{d}X_{jk}\right) _{d\times d}=\mathrm{d}(U\tilde{X%
}U^{-1})=U\left[ \mathrm{d}\tilde{X}+U^{-1}(\mathrm{d}U)\tilde{X}-\tilde{X}%
U^{-1}\mathrm{d}U\right] U^{-1},
\end{equation}%
where $\mathrm{d}U:=(\mathrm{d}U_{jk})_{d\times d}$. The line element in the
space of the entries of Hermitian matrices $X$ reads
\begin{equation}
\mathrm{d}s^{2}:=\text{tr}\left( \mathrm{d}X^{2}\right) =\sum_{j}(\mathrm{d}%
x_{j})^{2}-\sum_{k<l}(x_{k}-x_{l})^{2}\delta u_{kl}\delta u_{lk},
\label{line element}
\end{equation}%
where $\delta u_{kl}:=(U^{-1}\mathrm{d}U)_{kl}$. The flat Lebesgue measure $%
\mathbf{d}X$ can be induced from Eq. (\ref{line element}) as (see, e.g.,
Ref. \cite{sBook-RMT2})%
\begin{equation}
\mathbf{d}X=\varepsilon \prod_{j}\mathrm{d}x_{j}|\Delta (x)|^{2}\mathbf{d}%
\mu (U),  \label{GUE3}
\end{equation}%
where $|\Delta (x)|^{2}:=\prod_{k<l}(x_{k}-x_{l})^{2}$ is the squared
Vandermonde determinant, and $\mathbf{d}\mu (U)$ is the $\emph{uniform}$
probability (Haar) measure on the unitary group $\mathcal{U}(d)$ being
normalized $\int_{\mathcal{U}(d)}\mathbf{d}\mu (U)=1$ with a constant $%
\varepsilon $. Using Eq. (\ref{GUE2}) and Eq. (\ref{GUE3}) together with Eq.
(\ref{GUE1}), the GUE average reads%
\begin{eqnarray}
\langle f(X)\rangle _{\mathrm{GUE}} &=&\int f(X)\varrho _{\text{\textrm{GUE}}%
}(x_{1},\dots ,x_{d})\prod_{j}\mathrm{d}x_{j}\int_{\mathcal{U}(d)}\mathbf{d}%
\mu (U)  \notag \\
&=&\int f(X)\varrho _{\text{\textrm{GUE}}}(x_{1},\dots ,x_{d})\prod_{j}%
\mathrm{d}x_{j},  \label{X1}
\end{eqnarray}%
where $\varrho _{\text{\textrm{GUE}}}(x_{1},\dots ,x_{d}):=C^{\prime }\exp
(-\sum_{j}x_{j}^{2})|\Delta (x)|^{2}$ ($C^{\prime }=C\varepsilon $, given in
e.g. Ref. \cite{sBook-RMT}) is the the $d$-point joint probability
distribution of the eigenvalues, and the normalized condition of Haar
measure has been employed in the second line of Eq. (\ref{X1}).
Specifically, Equation (\ref{X1}) can be written as
\begin{equation}
\langle f(X)\rangle _{\mathrm{GUE}}=\int f(X)\varrho _{\mathrm{GUE}}(x)%
\mathrm{d}x,  \label{X2}
\end{equation}%
when we just consider the level density, i.e., $\varrho _{\mathrm{GUE}%
}(x)=\int \varrho _{\mathrm{GUE}}(x_{1},\dots ,x_{d})\prod_{j=2}\mathrm{d}%
x_{j}$.

Since the probability measure $\mathbf{d}\mu (X)$ is invariant under the
unitary conjugation of $X$ ($X\rightarrow UXU^{-1}$) \cite{sBook-Tao}, we
can employ the Haar measure to simplify the calculation. Equation (\ref{X1})
can be rewritten as

\begin{eqnarray}
\langle f(X)\rangle _{\mathrm{GUE}} &=&\int \varrho _{\mathrm{GUE}%
}(x_{1},\dots ,x_{d})\left[ \int_{\mathcal{U}(d)}f(UXU^{-1})\mathbf{d}\mu (U)%
\right] \prod_{j}\mathrm{d}x_{j}  \notag \\
&=&\int \left\langle f(X)\right\rangle _{\mathrm{Haar}}\text{\textrm{D}}x,
\label{X3}
\end{eqnarray}%
where we have introduced \textrm{D}$x:=\varrho _{\mathrm{GUE}}(x_{1},\dots
,x_{d})\prod_{j}\mathrm{d}x_{j}$, and the Haar average
\begin{equation}
\left\langle f(X)\right\rangle _{\mathrm{Haar}}:=\int_{\mathcal{U}%
(d)}f(UXU^{-1})\mathbf{d}\mu (U),
\end{equation}%
in the second line for simplicity.

Note that as long as $\varrho _{\text{\textrm{GUE}}}(x_{1},\dots ,x_{d})$ is
known, the \textrm{GUE} average can be evaluated using Eq. (\ref{X1}).
However, the calculation can be greatly simplified with Eq. (\ref{X3})
provided the moment function (\textrm{Haar} average) of the unitary group is
given (see e.g., the following section).

\subsubsection{\textit{2. Proof of Eq. (4) in the main text}}

\bigskip In this section, we give a proof of the following bound
\begin{equation}
D_{\text{\textrm{GUE}}}\leq \Gamma \frac{d^{2}}{d+1},  \label{up}
\end{equation}%
where $\Gamma :=\sum_{\mu }\gamma _{\mu }$, and $d$ is the Hilbert space
dimension of the system. The equality is achieved when the initial state is
pure [i.e., Eq. (4) in the main text].

\textit{Proof}. \ In this work, we consider the chaotic decoherence
channels, with Lindblad operators $\{V_{\mu }\}$ sampled from random the
Gaussian unitary ensemble $(\mathrm{GUE})$. Therefore, $\{V_{\mu }\}$ are
the Hermitian operators, and the decoherence rate [Eq. (2) in the main text]
can be written as%
\begin{equation}
D:=-\frac{2\text{tr}(\rho _{0}\dot{\rho}_{0})}{\text{tr}\left( \rho
_{0}^{2}\right) }=\frac{2\sum_{\mu }\gamma _{\mu }\widetilde{\mathrm{var}}%
_{\rho _{0}}(V_{\mu })}{\text{tr}\left( \rho _{0}^{2}\right) },
\label{DecRate}
\end{equation}%
where $\widetilde{\mathrm{var}}_{\rho _{0}}(X):=\left\langle \rho
_{0}X^{2}\right\rangle _{\rho _{0}}-\left\langle X\rho _{0}X\right\rangle
_{\rho _{0}}$, with $\left\langle \cdot \right\rangle _{\rho _{0}}:=$tr$%
\left( \rho _{0}\cdot \right) $, is the modified variance. For an arbitrary
\textit{fixed} initial state (meaning that $\rho _{0}$ is chosen independent
of $V_{\mu }$), the decoherence rate averaged over the GUE is given by
\begin{equation}
D_{\mathrm{GUE}}=\frac{2\sum_{\mu }\gamma _{\mu }\left[ \text{tr}\left( \rho
_{0}^{2}\left\langle V_{\mu }^{2}\right\rangle _{\mathrm{GUE}}\right) -\text{%
tr}\left( \rho _{0}\left\langle V_{\mu }\rho _{0}V_{\mu }\right\rangle _{%
\mathrm{GUE}}\right) \right] }{\text{tr}(\rho _{0}^{2})}.  \label{DGUE}
\end{equation}%
For simplicity, we assume that all decoherence channels $\{V_{\mu }\}$ are
independent and uncorrelated. Therefore, in the following proof, we will
drop the subscript \textquotedblleft $\mu $\textquotedblright\ temporarily
for clarity. \newline

According to Eq. (\ref{X3}), we have%
\begin{eqnarray}
\left\langle V^{2}\right\rangle _{\mathrm{GUE}} &=&\int \left\langle
V^{2}\right\rangle _{\mathrm{Haar}}\mathrm{D}v  \notag \\
&=&\int \left[ \int_{\mathcal{U}(d)}UV^{2}U^{-1}\mathbf{d}\mu (U)\right]
\mathrm{D}v  \notag \\
&=&\int \left[ \text{tr}(V^{2})\frac{\mathds{1}_{d}}{d}\right] \mathrm{D}v
\notag \\
&=&\text{tr}(\left\langle V^{2}\right\rangle _{\mathrm{GUE}})\frac{\mathds{1}%
_{d}}{d},
\end{eqnarray}%
where in the third line we have employed the \textit{second} moment function
of the unitary group \cite{sHaar}
\begin{equation}
\int_{\mathcal{U}(d)}UXU^{-1}\mathbf{d}\mu (U)=\text{tr}(X)\frac{\mathds{1}%
_{d}}{d}.
\end{equation}%
Then

\begin{equation}
\text{tr}\left( \rho _{0}^{2}\left\langle V^{2}\right\rangle _{\mathrm{GUE}%
}\right) =\frac{1}{d}\text{tr}(\left\langle V^{2}\right\rangle _{\mathrm{GUE}%
})\text{tr}(\rho _{0}^{2}).  \label{one}
\end{equation}

On the other hand,

\begin{eqnarray}
\left\langle V\rho _{0}V\right\rangle _{\mathrm{GUE}} &=&\int \left\langle
V\rho _{0}V\right\rangle _{\mathrm{Haar}}\mathrm{D}v  \notag \\
&=&\int \left[ \int_{\mathcal{U}(d)}UVU^{-1}\rho _{0}UVU^{-1}\mathbf{d}\mu
(U)\right] \mathrm{D}v  \notag \\
&=&\int \left[ \frac{d\text{tr}(V^{2})-\text{tr}(V)^{2}}{d(d^{2}-1)}%
\mathds{1}_{d}+\frac{d\text{tr}(V)^{2}-\text{tr}(V^{2})}{d(d^{2}-1)}\rho _{0}%
\right] \mathrm{D}v  \notag \\
&=&\frac{d\text{tr}(\left\langle V^{2}\right\rangle _{\mathrm{GUE}%
})-\left\langle \text{tr}(V)^{2}\right\rangle _{\mathrm{GUE}}}{d(d^{2}-1)}%
\mathds{1}_{d}+\frac{d\left\langle \text{tr}(V)^{2}\right\rangle _{\mathrm{%
GUE}}-\text{tr}(\left\langle V^{2}\right\rangle _{\mathrm{GUE}})}{d(d^{2}-1)}%
\rho _{0},
\end{eqnarray}%
where in the third line we have employed the \textit{fourth} moment function
of the unitary group \cite{sCollins,sBreuer13}
\begin{equation}
\int_{\mathcal{U}(d)}UX_{1}U^{-1}X_{2}UX_{3}U^{-1}\mathbf{d}\mu (U)=\frac{d%
\text{tr}(X_{1}X_{3})-\text{tr}(X_{1})\text{tr}(X_{3})}{d(d^{2}-1)}\text{tr}%
(X_{2})\mathds{1}_{d}+\frac{d\text{tr}(X_{1})\text{tr}(X_{3})-\text{tr}%
(X_{1}X_{3})}{d(d^{2}-1)}X_{2}.
\end{equation}%
It then follows that%
\begin{eqnarray}
\text{tr}(\rho _{0}\left\langle V\rho _{0}V\right\rangle _{\mathrm{GUE}}) &=&%
\frac{d\text{tr}(\left\langle V^{2}\right\rangle _{\mathrm{GUE}%
})-\left\langle \text{tr}(V)^{2}\right\rangle _{\mathrm{GUE}}}{d(d^{2}-1)}+%
\frac{d\left\langle \text{tr}(V)^{2}\right\rangle _{\mathrm{GUE}}-\text{tr}%
(\left\langle V^{2}\right\rangle _{\mathrm{GUE}})}{d(d^{2}-1)}\text{tr}%
\left( \rho _{0}^{2}\right)  \notag \\
&\geq &\frac{\text{tr}(\left\langle V^{2}\right\rangle _{\mathrm{GUE}%
})+\left\langle \text{tr}(V)^{2}\right\rangle _{\mathrm{GUE}}}{d(d+1)}\text{%
tr}\left( \rho _{0}^{2}\right) ,  \label{two}
\end{eqnarray}%
where the equality holds when $\rho _{0}$ is pure.

Substituting Eq. (\ref{one}) and Eq. (\ref{two}) (and recovering the
subscript \textquotedblleft $\mu $\textquotedblright\ of $V$) into Eq. (\ref%
{DGUE}), we have

\begin{eqnarray}
D_{\mathrm{GUE}} &\leq &\frac{2d}{d+1}\sum_{\mu }\gamma _{\mu }\left[
\left\langle \text{tr}\left( \rho _{\beta =0}V_{\mu }^{2}\right)
\right\rangle _{\mathrm{GUE}}-\left\langle \text{tr}\left( \rho _{\beta
=0}V_{\mu }\right) ^{2}\right\rangle _{\mathrm{GUE}}\right]  \notag \\
&=&\frac{2d}{d+1}\sum_{\mu }\gamma _{\mu }\left[ \left\langle \text{tr}%
\left( \rho _{\beta =0}V_{\mu }^{2}\right) \right\rangle _{\mathrm{GUE}}%
\right]  \notag \\
&=&\Gamma \frac{d^{2}}{d+1},  \label{De}
\end{eqnarray}%
where $\rho _{\beta =0}=\mathds{1}_{d}/d$ is the thermal state at infinite
temperature. In addition, we have used $\left\langle \text{tr}\left( \rho
_{\beta =0}V_{\mu }^{2}\right) \right\rangle _{\mathrm{GUE}}=d/2$ and $%
\left\langle \text{tr}\left( \rho _{\beta =0}V_{\mu }\right)
^{2}\right\rangle _{\mathrm{GUE}}=0$, both proven in \textit{Sec. \ref%
{SecD19}}, in the derivation of second and third lines in Eq. (\ref{De}).
\hfill $\square $

\begin{figure}[tbp]
\centering{}\includegraphics[width=3.15in]{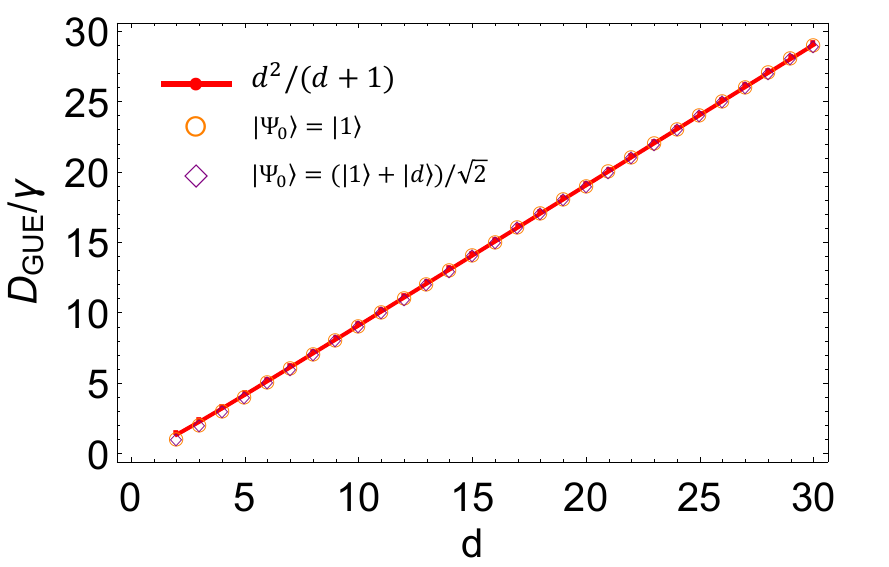}
\caption{Decoherence rates $D_{\mathrm{GUE}}/\protect\gamma $ versus the
dimension $d$ of chaotic decoherence channels sampled by random matrices.
Analytical formula derived by Haar measure (red dots) is in comparison with
numerical simulations over 20000 realizations of the GUE with fixed initial
states $\left\vert \Psi _{0}\right\rangle =\left\vert 1\right\rangle $
(orange circles), and $\left\vert \Psi _{0}\right\rangle =(\left\vert
1\right\rangle +\left\vert d\right\rangle )/\protect\sqrt{2}$ (purple
diamonds), respectively. }
\label{Fig_Haar}
\end{figure}

Note that the above proof is completely strict with no approximation and
valid for all dimensions. We can also employ the partition function method
to the calculation of $D_{\mathrm{GUE}}$, which will include some
approximations (see, e.g., in Section \ref{SecTFD} for initial thermofield
double states).

Equation (\ref{up}) provides an upper bound to the decoherence rate in GUE.
The equality is achieved when the initial state is a pure fixed state, which
is just the case we discuss in the main text. To verify Eq. (\ref{up}) as
well as Eq. (4) in the main text, in Fig. (\ref{Fig_Haar}), we choose the
one decoherence channel as an example (i.e., $\mu =1$, and denote $\gamma
_{\mu }=\gamma $) and compare the analytical expression with numerical
simulations over 20000 realizations of the GUE for two different initial
fixed pure states. In accordance with the proof, analytical results
accurately match the decoherence rate obtained from the numerical
simulations by averaging over different realizations of the Lindblad
operators.

\subsubsection{\textit{3. Derivation of Eq. (\protect\ref{De})}}

\label{SecD19}

\emph{1.---} Proof of $\left\langle \text{tr}\left( \rho _{\beta =0}V_{\mu
}^{2}\right) \right\rangle _{\mathrm{GUE}}=d/2$ in Eq. (\ref{De}).

\textit{Proof}. In the following, we drop the subscript \textquotedblleft $%
\mu $\textquotedblright\ temporarily for clarity

\begin{equation}
\left\langle \text{tr}\left( \rho _{\beta =0}V^{2}\right) \right\rangle _{%
\mathrm{GUE}}=\frac{1}{d}\left\langle \text{tr}\left( V^{2}\right)
\right\rangle _{\text{\textrm{GUE}}}=\frac{1}{d}\left[ \int v^{2}\varrho _{%
\mathrm{GUE}}(v)\mathrm{d}v\right] .  \label{a1}
\end{equation}%
The eigenvalue density averaged over \textrm{GUE} is given by \cite%
{sBook-RMT}%
\begin{equation}
\varrho _{\mathrm{GUE}}(v)=\sum_{l=0}^{d-1}\phi _{l}(v)^{2},\text{ and }\phi
_{l}(v):=\frac{e^{-\frac{v^{2}}{2}}\mathrm{H}_{l}(v)}{\sqrt{\sqrt{\pi }%
2^{l}l!}},  \label{a2}
\end{equation}%
where $\mathrm{H}_{l}(x)$ are the Hermite polynomials. Then we have%
\begin{equation}
\int v^{2}\varrho _{\mathrm{GUE}}(v)\mathrm{d}v=\frac{1}{\sqrt{\pi }}%
\sum_{l=0}^{d-1}\frac{1}{2^{l}l!}\int e^{-v^{2}}\left[ v\mathrm{H}_{l}(v)%
\right] ^{2}\mathrm{d}v=\frac{1}{2}\sum_{l=0}^{d-1}(2l+1)=\frac{d^{2}}{2},
\label{a3}
\end{equation}%
where we have employed the equality $x\mathrm{H}_{k}(x)=\frac{1}{2}\mathrm{H}%
_{k+1}(x)+k\mathrm{H}_{k-1}(x)$ and the orthogonality of the Hermite
polynomials. Substituting Eq. (\ref{a3}) into Eq. (\ref{a1}), we have $%
\left\langle \text{tr}\left( \rho _{\beta =0}V^{2}\right) \right\rangle _{%
\mathrm{GUE}}=d/2$. \hfill $\square $

\emph{2.---} Proof of $\left\langle \text{tr}\left( \rho _{\beta =0}V_{\mu
}\right) ^{2}\right\rangle _{\mathrm{GUE}}=0$ in Eq. (\ref{De}).

\textit{Proof}. As before, we drop the subscript \textquotedblleft $\mu $%
\textquotedblright\ temporarily for clarity

\begin{eqnarray}
\left\langle \text{tr}\left( \rho _{\beta =0}V\right) ^{2}\right\rangle _{%
\mathrm{GUE}} &=&\frac{1}{d^{2}}\left\langle \text{tr}\left( V\right)
^{2}\right\rangle _{\text{\textrm{GUE}}}  \notag \\
&=&\frac{1}{d^{2}}\left[ \int v^{2}\varrho _{\mathrm{GUE}}(v)\mathrm{d}v+%
\left[ \int v\varrho _{\mathrm{GUE}}(v)\mathrm{d}v\right] ^{2}+\int \int
vv^{\prime }\varrho _{\mathrm{GUE}}^{c}(v,v^{\prime })\mathrm{d}v\mathrm{d}%
v^{\prime }\right] ,  \label{b1}
\end{eqnarray}%
where the first item is just Eq. (\ref{a3}), and the rest items are from the
2-point correlation function $\varrho _{\mathrm{GUE}}(v,v^{\prime })=\varrho
_{\mathrm{GUE}}(v)\varrho _{\mathrm{GUE}}(v^{\prime })+\varrho _{\mathrm{GUE}%
}^{c}(v,v^{\prime })$, with
\begin{equation}
\varrho _{\mathrm{GUE}}^{c}(v,v^{\prime }):=-\sum_{k,l=0}^{d-1}\phi
_{k}(v)\phi _{l}(v)\phi _{k}(v^{\prime })\phi _{l}(v^{\prime })  \label{b2}
\end{equation}%
denoting the connected two-level correlation function \cite{sBook-RMT}. As
in the first proof, the second item in Eq. (\ref{b1}) reads

\begin{equation}
\int v\varrho _{\mathrm{GUE}}(v)\mathrm{d}v=0.  \label{b3}
\end{equation}%
In the following, we focus on the third item in Eq. (\ref{b1})

\begin{eqnarray}
\int \int vv^{\prime }\varrho _{\mathrm{GUE}}^{c}(v,v^{\prime })\mathrm{d}v%
\mathrm{d}v^{\prime } &=&-\sum_{k,l=0}^{d-1}\int v\phi _{k}(v)\phi _{l}(v)%
\mathrm{d}v\int v^{\prime }\phi _{k}(v^{\prime })\phi _{l}(v^{\prime })%
\mathrm{d}v^{\prime }  \notag \\
&=&-\frac{1}{\pi }\sum_{k,l=0}^{d-1}\frac{1}{2^{k}k!2^{l}l!}\int ve^{-\frac{%
v^{2}}{2}}\mathrm{H}_{k}(v)\mathrm{H}_{l}(v)\mathrm{d}v\int v^{\prime }e^{-%
\frac{v^{\prime 2}}{2}}\mathrm{H}_{k}(v^{\prime })\mathrm{H}_{l}(v^{\prime })%
\mathrm{d}v^{\prime }  \notag \\
&=&-\frac{1}{2}\sum_{k=0}^{d-1}(2k+1)=-\frac{d^{2}}{2},  \label{b4}
\end{eqnarray}%
where we have employed the equality%
\begin{equation}
\int xe^{-\frac{x^{2}}{2}}\mathrm{H}_{k}(x)\mathrm{H}_{l}(x)\mathrm{d}x=%
\frac{1}{2}\sqrt{\pi }(k+1)!2^{k+1}\delta _{l,k+1}+k\sqrt{\pi }%
(k-1)!2^{k-1}\delta _{l,k-1},  \label{b5}
\end{equation}%
which can be proved with the generating function $e^{2xt-t^{2}}=\sum_{k=0}^{%
\infty }\mathrm{H}_{k}(x)\frac{t^{k}}{k!}$ of Hermite polynomials. With Eq. (%
\ref{b3}), Eq. (\ref{b4}), and Eq. (\ref{a3}), we have $\left\langle \text{tr%
}\left( \rho _{\beta =0}V\right) ^{2}\right\rangle _{\mathrm{GUE}}\equiv 0$.
\hfill $\square $


\subsection{B. Proof of inequality $\mathrm{var}_{_{\protect\rho }}(X)\leq
\left\Vert X\right\Vert ^{2}$}

\textit{Proof}.%
\begin{equation}
\begin{split}
\mathrm{var}_{_{\rho }}(X)& =\text{tr}\left( \rho X^{2}\right) -\left[ \text{%
tr}\left( \rho X\right) \right] ^{2} \\
& \leq \text{tr}\left( \rho X^{2}\right) =\sum_{l}x_{l}^{2}\left\langle
l\right\vert \rho \left\vert l\right\rangle \\
& \leq \max \{x_{l}^{2}\}\sum_{l}\left\langle l\right\vert \rho \left\vert
l\right\rangle \\
& =x_{M}^{2}=\left\Vert X\right\Vert ^{2},
\end{split}%
\end{equation}%
where $x_{l}$ is the eigenvalue of $X$, $x_{M}$ is the maximum eigenvalue of
$\sqrt{X^{\dag }X}$, and $\left\Vert X\right\Vert $ is the spectral
norm.\hfill $\square $

\subsection{C. Master equations for the dynamics of the noise-averaged density
matrix}

\subsubsection{\textit{1. Stochastic fluctuating master equations}}

The dynamics of the noise-averaged density matrix $\rho _{t}=\left\langle
|\Psi _{t}^{\xi }\rangle \langle \Psi _{t}^{\xi }|\right\rangle _{\xi }$,
with $\xi :=\{\xi _{t}^{\mu }\}$ for simplicity, can be derived using
Novikov's theorem \cite{sNovikov,sChenu2017}. In this appendix we employ
another method, i.e., It\^{o} calculus, to derive the master equation. The
stochastic Schr\"{o}dinger equation of a quantum system $H_{0}$ perturbed by
the real Gaussian white noises $\hbar \sum_{\mu }\sqrt{\gamma _{\mu }}\xi
_{t}^{\mu }V_{\mu }$ is given by%
\begin{equation}
i\hbar \frac{\mathrm{d}|\Psi _{t}^{\xi }\rangle }{\mathrm{d}t}=\left(
H_{0}+\hbar \sum_{\mu }\sqrt{\gamma _{\mu }}\xi _{t}^{\mu }V_{\mu }\right)
|\Psi _{t}^{\xi }\rangle ,  \label{SF}
\end{equation}%
which, in the It\^{o} form, can be written as
\begin{equation}
\mathrm{d}|\Psi _{t}^{\xi }\rangle =-\frac{i}{\hbar }H_{0}\mathrm{d}t|\Psi
_{t}^{\xi }\rangle -i\sum_{\mu }\sqrt{\gamma _{\mu }}V_{\mu }\mathrm{d}%
W_{t}^{\mu }|\Psi _{t}^{\xi }\rangle -\frac{1}{2}\sum_{\mu }\gamma _{\mu
}V_{\mu }^{2}\mathrm{d}t|\Psi _{t}^{\xi }\rangle ,
\end{equation}%
where $\mathrm{d}W_{t}^{\mu }$, defined from $\xi _{t}^{\mu }:=\mathrm{d}%
W_{t}^{\mu }/\mathrm{d}t$, is an It\^{o} stochastic differential satisfying
the standard It\^{o} calculus rules, $\mathrm{d}W_{t}^{\mu }\mathrm{d}%
W_{t}^{\nu }=\delta _{\mu \nu }\mathrm{d}t$ and $\mathrm{d}W_{t}^{\mu }%
\mathrm{d}t=\mathrm{d}t^{2}=0$. According to the Leibnitz chain rule of It%
\^{o} calculus, $\mathrm{d}(XY)=(\mathrm{d}X)Y+X\mathrm{d}Y+(\mathrm{d}X)(%
\mathrm{d}Y)$ \cite{sSSE-PRD}, the corresponding Liouville-von Neumann
equation for the density matrix $\rho _{t}^{\xi }=|\Psi _{t}^{\xi }\rangle
\langle \Psi _{t}^{\xi }|$ is given by
\begin{equation}
\mathrm{d}\rho _{t}^{\xi }=-\frac{i}{\hbar }[H_{0},\rho _{t}^{\xi }]\mathrm{d%
}t-\frac{1}{2}\sum_{\mu }\gamma _{\mu }\left[ V_{\mu },\left[ V_{\mu },\rho
_{t}^{\xi }\right] \right] \mathrm{d}t-i\sum_{\mu }\sqrt{\gamma _{\mu }}%
\left[ V_{\mu },\rho _{t}^{\xi }\right] \mathrm{d}W_{t}^{\mu }.
\end{equation}%
Taking the stochastic expectation of the above equation, and considering $%
\left\langle XdW_{t}^{\mu }\right\rangle _{\xi }=0$ \cite{sSSE-PRD} gives
the evolution equation for $\rho _{t}=\langle \rho _{t}^{\xi }\rangle _{\xi
} $ as
\begin{equation}
\dot{\rho}_{t}=-\frac{i}{\hbar }[H_{0},\rho _{t}]-\frac{1}{2}\sum_{\mu
}\gamma _{\mu }\left[ V_{\mu },\left[ V_{\mu },\rho _{t}\right] \right] ,
\label{ME}
\end{equation}%
which corresponds to a master equation with Hermitian Lindblad operators.

\subsubsection{\textit{2. Examples}}

\textit{Example 1}.--Consider a general $k$-body long-range Ising model in a
transverse-field $h$ with the following Hamiltonian%
\begin{equation}
H_{0}=-\sum_{l_{1}<\cdots <l_{k}}^{n}J_{l_{1}<\cdots <l_{k}}\Lambda
_{l_{1}<\cdots <l_{k}}-h\sum_{l}^{n}\sigma _{l}^{x},
\end{equation}%
where $J_{l_{1}<\cdots <l_{k}}$ denote the coupling constants, $\Lambda
_{l_{1}<\cdots <l_{k}}:=\sigma _{l_{1}}^{z}\otimes \cdots \otimes \sigma
_{l_{k}}^{z}\otimes \mathds{1}_{\delta \neq l_{1},\cdots ,l_{k}}$, and $%
\sigma ^{\alpha }$ are the usual Pauli operators with $\alpha \in \{x,y,z\}$%
. By adding a single real Gaussian white noise to the coupling constants,
i.e.,
\begin{equation}
J_{l_{1}<\cdots <l_{k}}\rightarrow J_{l_{1}<\cdots <l_{k}}+\hbar \sqrt{%
\gamma }\xi _{t},
\end{equation}%
the noise-averaged density matrix obeys the master equation Eq. (\ref{ME}),
with a symmetric Lindblad operator
\begin{equation}
V=\sum_{l_{1}<\cdots <l_{k}}^{n}\Lambda _{l_{1}<\cdots <l_{k}}.
\end{equation}

Note that the same master equation arise from a variety of decoherence
sources. In the present example, the decoherence source is the stochastic
Gaussian white noise. In the main text, the Lindblad operator [i.e., Eq. (5)
in the main text] of the $k$-body long-range interactions is general,
independent of any specific decoherence sources.

\textit{Example 2}.--Two-body random ensembles (TBRE) has received
considerable attention in the theory of random matrices for many-body
quantum systems. Here, we consider an embedded ensemble with random two-body
interactions in a spin chain \cite{s2-bodyRE}
\begin{equation}
H_{0}=\sum_{l=1}^{n-1}\sum_{\alpha ,\alpha ^{\prime }=x,y,z}A_{l,\alpha
,\alpha ^{\prime }}\sigma _{l}^{\alpha }\sigma _{l+1}^{\alpha ^{\prime
}}+\sum_{l=1}^{n}\sum_{\alpha =x,y,z}B_{l,\alpha }\sigma _{l}^{\alpha },
\end{equation}%
where we have assumed the open boundary conditions, $A_{l,\alpha ,\alpha
^{\prime }}$ are the random two-body interaction variables, and $B_{l,\alpha
}$ denote the random external fields. Similar to \textit{Example 1}, a
single real Gaussian white noise is added to the random two-body interaction
variables
\begin{equation}
A_{l,\alpha ,\alpha ^{\prime }}\rightarrow A_{l,\alpha ,\alpha ^{\prime
}}+\hbar \sqrt{\gamma }\xi _{t},
\end{equation}%
the noise-averaged density matrix also obeys the master equation Eq. (\ref%
{ME}), with a symmetric Lindblad operator%
\begin{equation}
V=\sum_{l=1}^{n-1}\sum_{\alpha ,\alpha ^{\prime }=x,y,z}\sigma _{l}^{\alpha
}\sigma _{l+1}^{\alpha ^{\prime }}.
\end{equation}%
Then the decoherence rate will be bounded by $D\leq 2\gamma \left\Vert
V\right\Vert ^{2}\leq 162\gamma (n-1)^{2},$ where we have used $\left\Vert
\sigma _{l}^{\alpha }\sigma _{l+1}^{\alpha ^{\prime }}\right\Vert =1$.
Obviously, the decoherence rate of this two-body random ensembles model will
increase at most in a polynomial way.


\textit{Example 3.--}We consider a composite system describing two
non-interacting subsystems Hamiltonian
\begin{equation}
\tilde{H}_{0}=H\otimes \mathds{1}+\mathds{1}\otimes H,
\end{equation}%
that are independently perturbed by Gaussian real white noises $H\rightarrow
(1+\hbar \sqrt{\gamma }\xi _{t}^{L(R)})H$. As a result, the dynamic is
generated by the fluctuating (stochastic) Hamiltonian
\begin{equation}
\tilde{H}_{t}=H\otimes \mathds{1}+\mathds{1}\otimes H+\hbar \sqrt{\gamma }%
(\xi _{t}^{L}H\otimes \mathds{1}+\mathds{1}\otimes \xi _{t}^{R}H).
\end{equation}%
Assuming $\xi _{t}^{L}\neq \xi _{t}^{R}$, the noise-averaged density matrix
obeys the master equation Eq. (\ref{ME}), with $\mu \in \{1,2\}$ and the
following choice of the Lindblad operators%
\begin{equation}
\tilde{V}_{1}=H\otimes \mathds{1},\text{ and }\tilde{V}_{2}=\mathds{1}%
\otimes H.
\end{equation}%
Note that the above example concerns the setting in which we discussed the
decoherence of a thermofield double state case in the main text.

We also note that decoherence in the energy eigenbasis arises as well from
uncertainties in the measurement of time~\cite{sE1,sE2}, due to the
inability to physically determine the value of the ideal time parameter $t$
with arbitrary precision. If one is limited to a non-ideal clock, the
observed evolution in terms of a physical time parameter is effectively
non-unitary, satisfying Eq.~(\ref{ME}) with Lindblad operators $\widetilde{V}%
=H\otimes \mathds{1}+\mathds{1}\otimes H$, where the constant $\gamma $
depends on the clock precision. Here, two alternatives open up: either time
intervals can be determined with arbitrary precision, or the laws of physics
put fundamental constraints on it. The latter case has been proposed by a
combination of general relativity and quantum mechanics arguments \cite%
{sE1,sE2}, in which the loss of unitary is considered fundamental.

\subsection{D. Decoherence rate of the thermofield double (TFD) state in GUE}

\label{SecTFD}

\subsubsection{\textit{1. Decoherence dynamics of the TFD state}}

The analysis in the main text is focused on the decoherence time extracted
from the short-time asymptotics of the purity decay. Under dephasing, the
purity decay is monotonic as a function of time. Thus, the decoherence rates
provide a conservative estimate to the actual decay dynamics and the rate is
expected to decrease as a function of time. To analyze the complete dynamics
we consider the master equation governing the density matrix of the
composite system with stochastic Hamiltonian. For a single realization of
the noise, the dynamics is described by the Liouville-von Neumann equation
\begin{equation}
\dot{\rho}_{t}^{\mathrm{\xi }}=\frac{-i}{\hbar }[(H\otimes \mathds{1}+%
\mathds{1}\otimes H)+\hbar \sqrt{\gamma }(\xi _{t}^{L}H\otimes \mathds{1}+%
\mathds{1}\otimes \xi _{t}^{R}H),\rho _{t}^{\mathrm{\xi }}].
\end{equation}%
The dynamics of the average density matrix over many realizations of the
noise reads, from (\ref{ME}), [also see \textit{Example 2} in Section C]
\begin{equation}
\dot{\rho}_{t}=\frac{-i}{\hbar }[(H\otimes \mathds{1}+\mathds{1}\otimes
H),\rho _{t}]-\frac{\gamma }{2}[H\otimes \mathds{1},[H\otimes \mathds{1}%
,\rho _{t}]]-\frac{\gamma }{2}[\mathds{1}\otimes H,[\mathds{1}\otimes H,\rho
_{t}]],  \label{ME2}
\end{equation}%
with the initial state being given by
\begin{eqnarray}
\rho _{0} &=&|\Phi _{0}\rangle \langle \Phi _{0}|  \notag \\
&=&\frac{1}{Z(\beta )}\sum_{k,\ell }e^{-\frac{\beta }{2}(E_{k}+E_{\ell
})}|k\rangle |k\rangle \langle \ell |\langle \ell |,
\end{eqnarray}%
for a given operator $H$, where $E_{k(\ell )}$ are the corresponding
eigenvalues.

The operators in Eq. (\ref{ME2}) being in their diagonal basis, the
time-evolution of the density matrix can be obtained in a closed form as
\begin{equation}
\dot{\rho}_{kk,\ell \ell }=\frac{2}{i\hbar }(E_{k}-E_{\ell }){\rho }%
_{kk,\ell \ell }-\gamma (E_{k}-E_{\ell })^{2}{\rho }_{kk,\ell \ell }.
\end{equation}%
Thus, the exact time-dependent density matrix is given by
\begin{eqnarray}  \label{eq:SMevolution}
\rho _{t} &=&\sum_{k,\ell }{\rho }_{kk,\ell \ell }(t=0)e^{-i\frac{2t}{\hbar }%
(E_{k}-E_{\ell })-\gamma t (E_{k}-E_{\ell })^{2}}|k\rangle |k\rangle \langle
\ell |\langle \ell |  \notag \\
&=&\frac{1}{Z(\beta )}\sum_{k,\ell }e^{-\frac{\beta }{2}(E_{k}+E_{\ell
})}e^{-i\frac{2t}{\hbar }(E_{k}-E_{\ell })-\gamma t(E_{k}-E_{\ell
})^{2}}|k\rangle |k\rangle \langle \ell |\langle \ell |.
\end{eqnarray}

We note that the fixed-point of the evolution
\begin{equation}
\rho _{\infty }=\frac{1}{Z(\beta )}\sum_{k}e^{-\beta E_{k}}|k\rangle
|k\rangle \langle k|\langle k|,
\end{equation}%
is a separable state. Thus, as $t\rightarrow \infty $ the off-diagonal
elements (so-called coherences) of the density matrix decay to zero, showing
that entanglement is lost in the decoherence process.

The purity of the time-dependent density matrix decays, as the time of
evolution goes by, according to
\begin{equation}
P_{t}=\frac{1}{Z(\beta )^{2}}\sum_{k,\ell }e^{-\beta (E_{k}+E_{\ell
})-2\gamma t(E_{k}-E_{\ell })^{2}}.
\end{equation}%
At long-times, it saturates at the value
\begin{equation}
P_{\infty }=\frac{1}{Z(\beta )^{2}}\sum_{k}e^{-2\beta E_{k}}=\frac{Z(2\beta )%
}{Z(\beta )^{2}},
\end{equation}%
which is precisely the purity of a canonical thermal state. This long-time
asymptotic limit is shared by the unitary dynamics \cite%
{sDyer17,sdelCampo17prd}.

For arbitrary $t$, we make use of the Hubbard-Stratonovich transformation to
write
\begin{equation}
e^{-2t\gamma (E_{k}-E_{\ell })^{2}}=\sqrt{\frac{1}{8\pi \gamma t}}%
\int_{-\infty }^{\infty }e^{-\frac{y^{2}}{8\gamma t}}e^{-iy(E_{k}-E_{\ell })}%
\mathrm{d}y.
\end{equation}%
This yields the following expression for the purity
\begin{equation}
P_{t}=\sqrt{\frac{1}{8\pi \gamma t}}\int_{-\infty }^{\infty }e^{-\frac{y^{2}%
}{8\gamma t}}\left\vert \frac{Z(\beta -iy)}{Z(\beta )}\right\vert ^{2}%
\mathrm{d}y,
\end{equation}%
in terms of the analytic continuation of the partition function. The later
has been extensively studied as a characterization of the spectral
properties of quantum chaotic systems as well as a proxy for information
scrambling; see \cite{sDyer17,sdelCampo17prd} and references therein.

We are interested in the ensemble dynamics of the purity $P_{t}$ with $H\in $
\textrm{GUE}, i.e.,
\begin{equation}
\left\langle P_{t}\right\rangle _{\text{\textrm{GUE}}}=\sqrt{\frac{1}{8\pi
\gamma t}}\int_{-\infty }^{\infty }e^{-\frac{y^{2}}{8\gamma t}}\left\langle
\left\vert \frac{Z(\beta -iy)}{Z(\beta )}\right\vert ^{2}\right\rangle _{%
\text{\textrm{GUE}}}\mathrm{d}y.
\end{equation}%
We rely on the annealing approximation to simplify its computation
\begin{equation}
\left\langle P_{t}\right\rangle _{\text{\textrm{GUE}}}\dot{=}\frac{1}{%
\left\langle Z(\beta )^{2}\right\rangle _{\text{\textrm{GUE}}}}\sqrt{\frac{1%
}{8\pi \gamma t}}\int_{-\infty }^{\infty }e^{-\frac{y^{2}}{8\gamma t}%
}\left\langle \left\vert Z(\beta -iy)\right\vert ^{2}\right\rangle _{\text{%
\textrm{GUE}}}\mathrm{d}y,
\end{equation}%
the accuracy of which is well-established (see, e.g. Ref. \cite{sChenu18}).
Explicit expressions for both $\left\langle Z(\beta )^{2}\right\rangle _{%
\text{GUE}}$ and $\left\langle \left\vert Z(\beta -iy)\right\vert
^{2}\right\rangle _{\text{\textrm{GUE}}}$ for finite dimensional Hilbert
space dimension can be calculated with the polynomial method introduced in
\textit{Sec. \ref{SecD19}} (also see, e.g. Ref. \cite{sChenu18}).

\subsubsection{\textit{2. Derivation of decoherence rate of the TFD state ---
Eq. (13)}}

\begin{figure}[t]
\centering{}\includegraphics[width=7in]{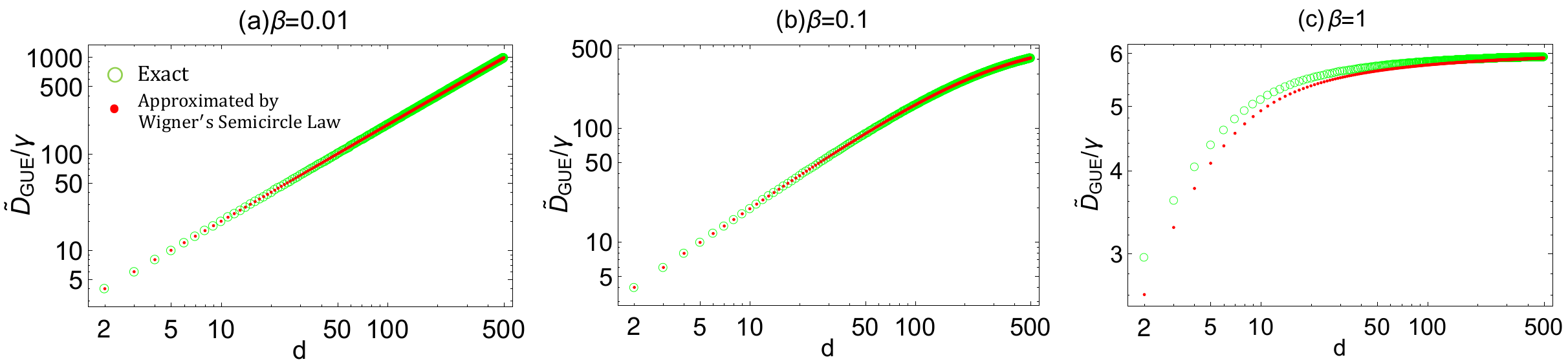}
\caption{Decoherence rates $\tilde{D}_{\mathrm{GUE}}/\protect\gamma $ versus
dimension $d$ with exact random matrices theory [Eq. (\protect\ref{E4})]
(green circles) and Wigner's semicircle approximations [Eq. (\protect\ref{F3}%
)] (red dots), displayed for (a) $\protect\beta =0.01$; (b) $\protect\beta %
=0.1$; and (c) $\protect\beta =1$, respectively.}
\label{Fig_WigSemi}
\end{figure}

The partition function of thermofield double states under the average over
GUE is given by
\begin{equation}
\left\langle Z(\beta )\right\rangle _{\text{\textrm{GUE}}}=\int \varrho _{%
\text{\textrm{GUE}}}(E)e^{-\beta E}\mathrm{d}v,  \label{E1}
\end{equation}%
with the averaged spectral density $\varrho _{\text{\textrm{GUE}}}(E)$
mentioned in Eq. (\ref{a2}) ($v\rightarrow $ $E$). By using the integration
of Hermite polynomials $\int \mathrm{d}xe^{-(x+a)^{2}}\mathrm{H}_{n}(x)%
\mathrm{H}_{m}(x)=\sqrt{\pi }2^{n}n!(-2a)^{m-n}\mathrm{L}%
_{n}^{(m-n)}(-2a^{2})$ for $n\leq m$ \cite{sBook-Int}, we have%
\begin{equation}
\begin{split}
\left\langle Z(\beta )\right\rangle _{\text{\textrm{GUE}}}& =\sum_{l=0}^{d-1}%
\frac{1}{2^{l}l!\sqrt{\pi }}e^{\frac{\beta ^{2}}{4}}\int e^{-(E+\frac{\beta
}{2})^{2}}\mathrm{H}_{l}(E)^{2}\mathrm{d}E\, \\
& =e^{\frac{\beta ^{2}}{4}}\sum_{l=0}^{d-1}\mathrm{L}_{l}\left( -\frac{\beta
^{2}}{2}\right) \\
& =e^{\frac{\beta ^{2}}{4}}\mathrm{L}_{d-1}^{(1)}\left( -\frac{\beta ^{2}}{2}%
\right) ,
\end{split}
\label{E3}
\end{equation}%
where $\mathrm{L}_{n}(x)$ are the Laguerre polynomials and $\mathrm{L}%
_{n}^{(\alpha )}(x)$ are the generalized Laguerre polynomials satisfying the
recurrence relation $\mathrm{L}_{n}^{(\alpha +1)}(x)=\sum_{l=0}^{n}\mathrm{L}%
_{l}^{(\alpha )}(x)$, with $\mathrm{L}_{n}^{(0)}(x)=\mathrm{L}_{n}(x)$. Then
the decoherence rate is given, from Eq. (11) in the main text and using the
annealing approximation (see, e.g., \textit{Sec. \ref{SecD3}}), by%
\begin{eqnarray}
\tilde{D}_{\text{\textrm{GUE}}} &\dot{=}&4\gamma \frac{\mathrm{d}^{2}}{%
\mathrm{d}\beta ^{2}}\ln \left\langle Z(\beta )\right\rangle _{\text{\textrm{%
GUE}}}  \notag \\
&=&2\gamma \Big[1+2\mathrm{F}_{1}^{(2)}\left( {-}\beta ^{2}/2\right) -2\beta
^{2}\left( \mathrm{F}_{1}^{(2)}\big({-}\beta ^{2}/2\big)\right) ^{2}+2\beta
^{2}\mathrm{F}_{1}^{(3)}\left( {-}\beta ^{2}/2\right) \Big],\text{ }
\label{E4}
\end{eqnarray}%
with \textrm{F}$_{l}^{(m)}(x):=\mathrm{L}_{d-m}^{(m)}(x)/\mathrm{L}%
_{d-l}^{(l)}(x)$.

In the large dimension case, the eigenvalue density of $E$ over Gaussian
random matrices average obeys the Wigner's semicircle law%
\begin{equation}
\varrho _{\text{\textrm{GUE}}}(E)=\frac{\sqrt{2d}}{\pi }\sqrt{1-\left( \frac{%
E}{\sqrt{2d}}\right) ^{2}},  \label{F1}
\end{equation}%
with $E\in \lbrack -\sqrt{2d},\sqrt{2d}]$ \cite{sBook-RMT}. Then, the
partition function Eq. (\ref{E3}) is given by \cite{sdelCampo17prd}%
\begin{equation}
\left\langle Z(\beta )\right\rangle _{\text{\textrm{GUE}}}=\frac{\sqrt{2d}%
\mathrm{I}_{1}(\sqrt{2d}\beta )}{\beta },  \label{F2}
\end{equation}%
where $\mathrm{I}_{n}(x)$ is the modified Bessel function of first kind and
order $n$. Using the fact that $I_{0}(x)=I_{2}(x)+(2/x)I_{1}(x)$, this leads
to%
\begin{equation}
\tilde{D}_{\mathrm{GUE}}=8\gamma d\left[ 1-\frac{3}{\sqrt{2d}\beta }g(\sqrt{%
2d}\beta )-g(\sqrt{2d}\beta )^{2}\right] ,  \label{F3}
\end{equation}%
with $g(x):=\mathrm{I}_{2}(x)/\mathrm{I}_{1}(x)$. Equation (\ref{F3})
corresponds to Eq. (11) in the main text.

According to Wigner's semicircle law, when the dimension is large, Eq. (\ref%
{F3}) matches Eq. (\ref{E4}) perfectly. In addition, numerical simulations
show that when $\beta \ll 1$, Eq. (\ref{F3}) also well agrees with Eq. (\ref%
{E4}), even when the dimension is not large [shown in Fig. \ref{Fig_WigSemi}%
].

\subsubsection{\textit{3. Annealing approximation}}

\label{SecD3}

In this section, we provide a brief instruction for the annealing
approximation used in Eq. (\ref{E4}). To simplify the calculations, we make
use of the annealed average over logarithm of the partition function $\ln
Z(\beta )$, i.e.,
\begin{equation}
\left\langle \ln [Z(\beta )]\right\rangle _{\text{\textrm{GUE}}}\doteq \ln
\left\langle Z(\beta )\right\rangle _{\text{\textrm{GUE}}}.  \label{D1}
\end{equation}%
In fact, according to Jensen's inequality $\left\langle \ln [Z(\beta
)]\right\rangle _{\text{\textrm{GUE}}}\leq \ln \left\langle Z(\beta
)\right\rangle _{\text{\textrm{GUE}}}$, since $\ln (x)$ is a concave
function. The equality is well satisfied in high dimensional systems (when
the dimension is not large, the annealing approximation is still valid in
high temperature regime), as verified by numerical simulations; see Fig. \ref%
{Fig_annealing}.

\begin{figure}[t]
\centering{}\includegraphics[width=7in]{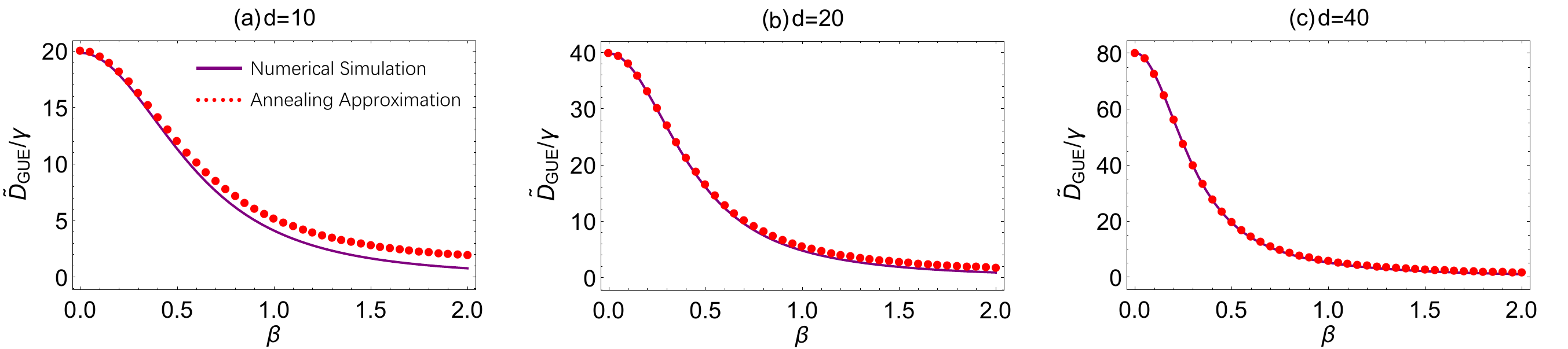}
\caption{Decoherence rates $\tilde{D}_{\mathrm{GUE}}/\protect\gamma $ versus
$\protect\beta $ of thermofield double states in GUE with exact numerical
simulations (solid line) and annealing approximations (red dots)
respectively. The numerical average is performed over 2000 realizations of
the GUE with the dimension (a) $d=10$; (b) $d=20$; and (c) $d=40$,
respectively.}
\label{Fig_annealing}
\end{figure}


\begin{thebibliography}{99}
\bibitem{note} Note that Eq. (\ref{DR}) may also include the dissipative
effect, however, such dissipation rates are negligible compared with the
decoherence rates since the relaxation timescale is many orders of longer
than the decoherence timescale \cite{Book-Decoherence}.
\end{thebibliography}
\end{document}